\documentclass[12pt,a4paper]{article}
\usepackage{jheppub}
\usepackage[T1]{fontenc} 
\usepackage{epsfig} 
\usepackage{amscd}
\usepackage{verbatim}
\usepackage{latexsym}
\usepackage{amsfonts,amsthm,amsmath,mathtools}

\newcommand{\tr}{{\rm tr}}

\notoc 

\title{\centering
A journey to $3d$: exact relations for adjoint SQCD from dimensional reduction\\
}
\author[a]{Antonio Amariti,}
\author[b,c]{Claudius Klare,}

\affiliation[a]{Laboratoire de Physique Th\'eorique de l'\'Ecole Normale Sup\'erieure \\
24 Rue Lhomond, Paris 75005, France}
\affiliation[b]{Institut de Physique Th\'eorique, CEA/Saclay \\
CNRS URA 2306, 
F-91191 Gif-sur-Yvette, France} 
\affiliation[c]{Institut des Hautes \'Etudes Scientifiques\\
35 Route de Chartres, F-91440 Bures-sur-Yvette, France}
               
\emailAdd{amariti@phys.lpt.ens}
\emailAdd{claudius.klare@cea.fr}

\abstract{In this note we elaborate on the reduction of four dimensional Seiberg duality with adjoint matter to three dimensions.
  We use the exact formulation of the superconformal index and of the partition function as instruments to test this reduction.
  We translate the identity between indices of the dual $4d$ theories to $3d$.
  This produces various new identities between partition functions of $3d$ dual phases.
  \vspace{6cm}
}

\begin{document}

\maketitle

\section{Introduction}

A general program in the study of supersymmetric gauge theories is the
connection between dualities in various dimensions.  A systematic
approach to the case with four supercharges in four and three
dimensions has been recently proposed in \cite{Aharony:2013dha}.

The connection is somewhat subtle, it turns out that a plain
dimensional reduction is too naive.  This fact
can be understood considering that in three dimensions there are
extra, axial, symmetries, whose $4d$ parents would be anomalous.  A
careful analysis requires keeping the $4d$ theory on a circle with
finite radius, whereby a superpotential breaking the axial symmetry is
generated.  This mechanism generates a well-defined, new, IR duality in
three dimensions from a four dimensional parent.  By adding real
masses such $3d$ dual pairs flow to more canonical dualities.  Along
the flow the superpotential disappears and the axial symmetry emerges.

This stepwise reduction has been performed for four dimensional SQCD
in \cite{Aharony:2013dha,Aharony:2013kma}, reproducing the dualities
of \cite{Aharony:1997gp,Karch:1997ux,Giveon:2008zn,Benini:2011mf,Aharony:2011ci}.
There is an extension of Seiberg duality for SQCD by adding an adjoint multiplet studied by Kutasov, Schwimmer and Seiberg (KSS) in
\cite{Kutasov:1995ve,Kutasov:1995np,Kutasov:1995ss}.  Recently
\cite{Nii:2014jsa} the above technology has been applied to this
duality reproducing the results of \cite{Kim:2013cma}\footnote{See
  \cite{Csaki:2014cwa} for another application of the stepwise
  reduction to the case of s-confining theories.}.

Exact results in supersymmetry, as from the recent progress in
localisation, lead to efficient tests of dualities.  For example, the
superconformal index on $S^3 \times S^1$
\cite{Romelsberger:2005eg,Kinney:2005ej} has been matched between
Seiberg dual phases
\cite{Dolan:2008qi,Spiridonov:2008zr,Spiridonov:2009za}.  A similar
matching has been done in \cite{Willett:2011gp,Benini:2011mf} for the
partition functions on the (squashed) $S^3$
\cite{Jafferis:2010un,Hama:2010av,Hama:2011ea}.  In both cases the
identification is equivalent to non-trivial integral identities
involving elliptic and hyperbolic Gamma functions respectively,
recently analysed in the mathematical literature
\cite{Rains15,Spiridonov15,VdB}.

Moreover, the integral expressions for the index and the partition
function provide a powerful tool for supporting the reduction set-up
described above.  It is possible to follow the reduction and the mass
deformation on the integral expressions.  This has been done in
\cite{Aharony:2013dha} for SQCD where the relations of \cite{VdB} have
been recovered from the ones of \cite{Rains15}.

In this paper, we study the dimensional reduction of SQCD with an
adjoint \cite{Nii:2014jsa} at the level of the index and the partition function.
Thereby we generate new integral identities involving
hyperbolic Gamma functions, generalising some of the results in \cite{VdB}.

The paper is organised as follows.
In section \ref{sec:review} we recap the electric-magnetic dualities discussed in the paper.
We also give a brief review of the mechanism relating them by dimensional reduction.
In section \ref{sec:I-mathcal-Z} we implement the reduction in the matrix integrals of the $4$ and $3$ dimensional theories,
deriving the integral identities.
In section \ref{sec:conclusioni} we conclude with some comments.
We added an appendix \ref{app:Z-3d} with few details about the partition function on the squashed three sphere.

\section{Dualities and dimensional reduction}
\label{sec:review}

In sections \ref{sec:KSS} and \ref{sec:KP} we review the four and three dimensional dualities with adjoint matter that are of interest in this paper.
In section \ref{sec:4d3d} we then review the dimensional reduction connecting them.

\subsection{$4d$ duality with adjoint matter}
\label{sec:KSS}
KSS duality \cite{Kutasov:1995ve,Kutasov:1995np,Kutasov:1995ss} 
involves four dimensional SQCD theories with a chiral multiplet in the adjoint representation.

The \emph{electric} phase is a  $SU(N_c)$ gauge theory with $N_f$ (anti-)fundamentals $Q$ and $\tilde Q$ and one adjoint $X$.
There is a superpotential 
\begin{equation}
\label{eq:W-el-adj}
W = \tr \, X^{k+1}
\end{equation}
with $k<N_c$.
The global symmetry group is $SU(N_f)_L \times SU(N_f)_R \times U(1)_B \times U(1)_R$,
we denote the $R$ charges of the (anti-)quarks as $R_Q$. 
The $R$ charge of the adjoint is fixed by $\eqref{eq:W-el-adj}$ as $R_X=2/(k+1)$.
The mesonic operators are
\begin{equation}
  \label{eq:mesons-KSS}
M_j = Q X^{j} \tilde Q \, , \quad \quad \quad j=0,\dots,k-1
\end{equation}
The theory has stable vacua if
$N_f > \frac{N_c}{k}$.

The \emph{magnetic} phase is an $SU(k N_f-N_c)$ gauge theory, with $N_f$ dual (anti-) fundamentals $q$ and $\tilde q$ and one adjoint $Y$.
There is also a set of $k$ gauge singlets, which correspond to the mesons $\eqref{eq:mesons-KSS}$ of the electric phase.
The superpotential is 
\begin{equation}
\label{magn4d}
W = \tr \, Y^{k+1} + \sum_{j=0}^{k-1} M_j q Y^{k-1-j} \tilde q 
\end{equation}
The $R$ charges of the dual (anti-)quarks are $R_q = R_X - R_Q$.

By gauging the non-anomalous global $U(1)_B$ in the electric and in the 
magnetic phase one can extend the duality to the unitary gauge groups $U(N_c)$ and $U(k N_f-N_c)$.
This is the case considered in this paper.

\subsection{$3d$ dualities with adjoint matter}
\label{sec:KP}
A similar duality exists in three dimensions.
It generalises the duality studied in \cite{Aharony:1997gp} by the inclusion of an adjoint
and has been discussed by Kim and Park in \cite{Kim:2013cma}.

The \emph{electric} phase is a $U(N_c)$ YM gauge theory with $N_f$ (anti-)fundamentals $Q$ and $\tilde Q$ and one adjoint $X$.
The superpotential is $\eqref{eq:W-el-adj}$, as in the $4d$ case.
The global symmetry group is 
 $SU(N_f)_L \times SU(N_f)_R \times U(1)_A \times U(1)_R\times U(1)_J$.
Recall that in three dimensions there is a topologically conserved current $J = * d F$,
where $F$ is the field strength of the $U(1)$ inside the $U(N)$ gauge group.
The charges of the chiral fields under the global symmetries are
v
There is a $2k$ dimensional unlifted Coulomb branch, parametrised by $t_{j, \pm}$, where $j=0,\dots,k-1$.
These coordinates have an UV interpretation as monopole operators $t_{0,\pm}$, dressed with powers of the adjoint
\begin{equation} \label{eq:mono-op} 
 t_{j,\pm} = \tr \, (t_{0,\pm}X^j)
\end{equation} 
The operators $t_{0,\pm}$ correspond to an excitation of the magnetic flux $(\pm 1, 0 , \dots , 0)$.

The \emph{magnetic} phase is a $U(k N_f-N_c)$ YM gauge theory with $N_f$ dual (anti-) fundamentals $q$ and $\tilde q$ and one adjoint $Y$.
As in the $4d$ case there are $k$ gauge singlets corresponding to the electric mesons $M_j$.
However in $3d$ we have additional $2k$ gauge singlets, $t_{j, \pm}$,
corresponding to the electric monopole operators $\eqref{eq:mono-op}$.
They couple to the magnetic theory through the superpotential
\begin{equation}
  \label{eq:W-KP-mag}
W = \tr \, Y^{k+1} + \sum_{j=0}^{k-1} M_j q Y^{k-j-1} \widetilde q
+ \sum_{j=0}^{k-1} \left(t_{j,+} \widetilde {t}_{k-1-j,-}
+ t_{j,-} \widetilde{t}_{k-1-j,+}\right)
\end{equation}
where the $2k$ ``monopole operators'' $\widetilde{t}_{j,\pm}$ are the Coulomb branch coordinates of the magnetic theory.
In the UV theory they correspond to monopole operators dressed with powers of the adjoint, $\tilde t_{j,\pm} = \tr \, (t_{0,\pm} Y^j)$.
The chiral fields and their charges under the global symmetries are

\begin{equation} 
\begin{array}{c||ccccc}
             &SU(N_f)_L & SU(N_f)_R    & U(1)_A & U(1)_R & U(1)_J \\
\hline
Q            &   N_f    &   1         &   1    & \Delta_Q &  0     \\
\widetilde Q  &   1    &\overline{N_f} &  1    & \Delta_Q &  0     \\
X &1&1&0&\frac{2}{k+1}&0
\end{array}
\end{equation}
\begin{equation} 
\begin{array}{c||ccccc}
             &SU(N_f)_L & SU(N_f)_R    & U(1)_A & U(1)_R & U(1)_J \\
\hline
q            &   \overline{N_f}    &   1  &   -1    & \frac{2}{k+1}-\Delta_Q &  0     \\
\widetilde q  &  1    & N_f & - 1    & \frac{2}{k+1}-\Delta_Q &  0     \\
Y &1&1&0&\frac{2}{k+1}&0\\
M_j&N_f&\overline{N_f}&2&2 \Delta_Q+\frac{2j}{k+1}&0\\
t_{j,\pm} &1 &1 &-N_f &N_f(1-\Delta_Q)-\frac{2}{k+1}(N_c-1-j) &\pm 1 \\
\widetilde{t}_{j,\pm} &1 &1 &N_f &N_f(\Delta_Q-1)+\frac{2}{k+1}(N_c+1+j) & \pm 1 
\end{array}
\end{equation}

\paragraph{Theories with CS terms.}
There is another $3d$ duality with adjoint matter, having Chern-Simons (CS) interactions \cite{Niarchos:2008jb,Niarchos:2009aa}.
The electric phase is a $U(N_c)_K$ gauge theory (where $K$ is the CS level). 
The magnetic phase has gauge group $U(k N_f-N_c +|K|)_{-K}$.
The Coulomb branch is lifted and no monopole operators appear in the description of the low energy theory.

\subsection{From four to three dimensional dualities}
\label{sec:4d3d}
In this section we review how one can obtain three dimensional from four dimensional dualities \cite{Aharony:2013dha}.
It turns out that a naive dimensional reduction on $\mathbb{R}^3 \times S^1$ does not lead to the right result.
This can be understood as a consequence of Seiberg duality being an IR property,
where the IR limit does not commute with the limit of the shrinking circle. 
More precisely, Seiberg duality is valid for energies much below the electric and the magnetic strong coupling scales,
$E \ll \Lambda^e, \Lambda^m$, which are each related to their coupling as $ \Lambda^b \sim e ^ {-1/g_4^2}$. 
Here $b$ is the coefficient of the $4d$ $\beta$-function.
When dimensionally reducing the theory on the circle $S^1$ with radius $r$ the coupling scales as $g_4^2 \sim r g_3^2$ which implies that $\Lambda \rightarrow 0$.
Hence the IR limit $E \ll \Lambda$ is not well defined in three dimensions and in that sense the two limits do not commute.
This argument mirrors the one presented in the introduction
where we mentioned the existence of axial symmetries in $3d$, whose $4d$ counterparts would be anomalous.
We are hence looking for a reduction which commutes with the low-energy limit of Seiberg duality and forbids the existence of the axial symmetry.

In order to obtain well-defined three dimensional dualities from four dimensional parents
we leave the theories on a circle $S^1$ with a \emph{finite} radius $r$.
At energies $E \ll 1/r$ that describes effectively three dimensional dynamics.  

The finite radius implies that the Coulomb branch is compact. 
Recall that the adjoint scalar $\sigma$ of the $3d$ vector multiplet comes from the forth component of the $4d$ gauge field, $\sigma = A_4$.
To make $\int_{S^1} A_4$ gauge invariant, one needs to require the periodicity $\sigma  \sim \sigma + 1/r$.
As a consequence of this compactness, there is an additional Affleck-Harvey-Witten (AHW) superpotential for the unlifted coordinates of the Coulomb branch\footnote{Recall
that the Coulomb branch for pure SYM is completely lifted, whenever two coordinates approach each other the gauge group is enhanced to $SU(2)$ and instantons
generate an AHW superpotential as in \cite{Affleck:1982as}.
For theories with matter fields charged under the gauge group, some directions of the Coulomb branch remain unlifted.
Similarly, when the Coulomb branch is \emph{compact}, additional $SU(2)$ enhancements can take place leading to the additional $W_{\mathbb{R}^3 \times S^1}$.},
for $U(N)$ theories generically looking like \cite{Seiberg:1996nz}
\begin{equation} 
  \label{eq:eta-superpot} 
  W_{\mathbb{R}^3 \times S^1} = \eta \, Y_+ Y_- \quad \quad
\end{equation} 
Here $\eta \sim e^{-1/(r g_3)^2}$ and $Y_{\pm}$ denote the unlifted directions of the Coulomb branch, with respective $U(1)_J$ charge $\pm 1$.
From the perspective of the effective $3d$ field theory the $Y$'s are just the low-energy coordinates parametrising the Coulomb branch.
One can though embed the $\eta$ deformed theory in a purely $3d$ UV completion, where the high-energy operators $Y_{\pm}$ typically correspond to monopole operators. 

Let us point out that $\eqref{eq:eta-superpot}$ breaks the axial symmetry.
Since this was the symmetry without $4d$ counterpart the construction indeed overcomes the problem of the mismatching symmetries. 
In order to obtain conventional $3d$ theories with the axial symmetry,
one can switch on a deformation by large real masses for some of the matter fields,
generating a flow to theories without superpotential $\eqref{eq:eta-superpot}$.

\subsubsection*{Reducing KSS to $3d$}
By generalising this analysis one can obtain the theories in section \ref{sec:KP} from the ones in section \ref{sec:KSS}.
In the presence of an adjoint with superpotential $\eqref{eq:W-el-adj}$ the Coulomb branch of a $U(N)$ theory has $2k$ unlifted directions, parametrised by $t_{i,\pm}$.
The $\eta$ superpotential in the electric theory is \cite{Nii:2014jsa}
\begin{equation} 
  \label{eq:eta-superpot-adjoint} 
  W_{\mathbb{R}^3 \times S^1} = \eta \sum_{j=0}^{k-1} t_{j,+} t_{k-1-j,-}
\end{equation} 
With an analogue superpotential in the magnetic phase we end up with a pair of dual three dimensional theories.

\subsubsection*{Flowing to the Kim-Park duality}
One can obtain Kim-Park duality by deforming the theories with $\eta$ superpotential.
As in \cite{Nii:2014jsa} we consider $N_c$ colours and $N_f+2$ flavours, where in the electric theory we assign large real masses 
\begin{equation} \label{eq:large-real-masses} 
  m_{Q_{N_f+1}} = -m_{Q_{N_f+2}} = - m_{\widetilde Q_{N_f+1}} = m_{\widetilde Q_{N_f+2}} = M
\end{equation} 
In the large $M$ limit this theory flows to the electric phase of Kim-Park duality.

The magnetic phase has $k(N_f+2)-N_c$ colours and $N_f+2$ flavours.
The real masses for the dual quarks follow from the electric ones, according to the global symmetries.
In \cite{Nii:2014jsa} the theory is perturbed by a polynomial superpotential in the adjoint.
This breaks the gauge group to $k$ sectors $U(n_i)$ with $\sum_i n_i = k(N_f +2)-N_c$.
In order to preserve the duality there is a non trivial vacuum structure for the scalar $\sigma$ in each $U(n_i)$ vector multiplet,
further breaking the gauge symmetry into $U(n_i-2) \times U(1)^2$.
The $2k$ $U(1)$ subsectors can be dualised to $2k$ $XYZ$ models \cite{Aharony:1997bx}, made of gauge singlets.
These singlets interact through a superpotential with the monopoles of the $U(n_i-2)$ sectors.
By turning off the polynomial deformation in the adjoint while taking the large $M$ limit,
one finds the magnetic phase of Kim-Park duality, with gauge group $U(k N_f -N_c)$ and superpotential $\eqref{eq:W-KP-mag}$.
The $2k$ surviving gauge singlets of the $XYZ$ models are identified with the electric monopoles.

In section \ref{sec:I-mathcal-Z} we reproduce this reduction mechanism on the index and the partition function. 
Actually, when restoring the dual gauge symmetry $U(\sum n_i)$ the $k$ $U(1)^2$ sectors become enhanced to $U(k)^2$,
each with one flavour and one adjoint.
In \cite{Nii:2014jsa} it is assumed that this enhancement does not affect the duality.
We support this assumption by using a duality discussed in \cite{Amariti:2014lla}.

\section{Superconformal index and partition function}
\label{sec:I-mathcal-Z}

In four dimensions a powerful check of Seiberg duality comes from matching the index on $S^3 \times S^1$ 
between the electric and the magnetic phases.
The matching involves non-trivial integral identities as shown in \cite{Dolan:2008qi,Spiridonov:2008zr,Spiridonov:2009za}.
A similar test exists for three dimensional $\mathcal{N}=2$ dualities \cite{Willett:2011gp,Benini:2011mf},
involving the partition function on the squashed $S^3$ \cite{Hama:2011ea}.
It is known that one can obtain the partition function from the index in the limit of a shrinking circle 
\cite{Dolan:2011rp,Gadde:2011ia,Imamura:2011uw,Niarchos:2012ah,Agarwal:2012wd}.

In this section we will check the mechanism for obtaining $3d$ dualities with an adjoint
from KSS duality at the level of these mathematical quantities.
We start from the integral identity between the indices of the KSS pair in $4d$.
By dimensional reduction we obtain an identity for the partition functions of the $3d$ pair with $\eta$ superpotential.
The identity for Kim-Park duality is then obtain by implementing the real mass deformation $\eqref{eq:large-real-masses}$.
Eventually we flow to the duality of \cite{Niarchos:2008jb,Niarchos:2009aa} with CS terms.

\subsection{From KSS duality to $3d$}

Let us consider the duality reviewed in section \ref{sec:KSS}.
The electric phase is a $U(N_c)$ gauge theory with one adjoint and $N_f+2$ (anti-)fundamental flavours.
The partition function on $S^3 \times S^1$ is
\begin{equation} \label{eq:index-el} 
  \begin{aligned}
    I_{el}= &\frac{(p;p)^{N_c}(q;q)^{N_c}}{N_c!}
\Gamma_e \big( (pq)^{\frac{1}{k+1}} \big)^{N_c}
\int \prod_{i=1}^{N_c} \frac{d z_i}{2 \pi i z_i}
\prod_{i< j} \frac{\Gamma_e \big((pq)^{\frac{1}{k+1}}(z_i/z_j)^{\pm 1}\big)}{ \Gamma_e\big( (z_i/z_j)^{\pm 1} \big)}
\\
\times &
\prod_{a,b=1}^{N_f+2} \prod_{i=1}^{N_c}
\Gamma_e \big(
(pq)^{\frac{R_Q}{2}} s_a z_i\big)
\Gamma_e \big(
(pq)^{\frac{R_Q}{2}} t_b^{-1} z_i^{-1}
\big)
  \end{aligned}
\end{equation}
where 
\begin{align}
(x;p) = \prod_{j=0}^{\infty} \big(1-x p^j\big)
&&
\Gamma_e(z) = \prod_{j,k=0}^{\infty} \frac{1-z^{-1}p^{j+1}q^{k+1}}{1-z p^j q^k}
\end{align}
The function $\Gamma_e$ is the elliptic Gamma function \cite{Rains15,Spiridonov15}. 
We also have defined $\Gamma_e( (z)^{\pm 1}) \equiv \Gamma_e(z) \Gamma_e (1/z)$.
The parameters $p$ and $q$ are the chemical potentials of the superconformal algebra,
$z_i$ the ones of the $U(N_c)$ gauge symmetry 
and $s_a$ and $t_a$ the ones of $SU(N_f)_L$ and $SU(N_f)_R$ respectively.

The magnetic phase is a $U(k(N_f+2)-N_c)\equiv U(\tilde N_c+2 k)$ gauge theory with one adjoint,
$N_f+2$ (anti-)fundamental flavours and the $k$ electric mesons.
The index is
\begin{eqnarray}
  \label{eq:index-mag}
  &&I_{mag}= \frac{(p;p)^{\tilde N_c+2k}(q;q)^{\tilde N_c+2k}}{(\tilde N_c+2k)!}
\Gamma_e\big( (pq)^{\frac{1}{k+1}} \big)^{\tilde N_c +2k}
\prod_{j=0}^{k-1} \prod_{a,b=1}^{N_f+2}
\Gamma_e\big((pq)^{\frac{j}{k+1}+R_Q} s_a t_b^{-1}\big)
\nonumber \\
&&
\int \prod_{i=1}^{\tilde N_c+2k} \frac{d z_i}{2 \pi i z_i}
\prod_{i<j} \frac{\Gamma_e\big((pq)^{\frac{1}{k+1}}(z_i/z_j)^{\pm 1}\big)}{
\Gamma_e\big( (z_i/z_j)^{\pm 1}\big)}
\prod_{a,b=1}^{N_f+2} \prod_{i=1}^{\tilde N_c+2k}
\Gamma_e\big(
(pq)^{\frac{R_q}{2}} t_a^{-1} z_i\big)
\Gamma_e\big(
(pq)^{\frac{R_q}{2}} s_b z_i^{-1}
\big) \nonumber \\
\end{eqnarray}
KSS duality predicts the integral identity $I_{el} = I_{mag}$. 
In \cite{Spiridonov:2009za} this identity has passed some partial nontrivial checks.
Note that for the identity to hold, the chemical potentials in the index have to satisfy the balancing condition
\begin{equation}
\label{eq:bal-cond-4d}
\prod_{a=1}^{N_f+2} s_a t_a^{-1} = (pq)^{N_f+2-2N_c/(k+1)}
\end{equation}

In order to reduce the four dimensional index to the three dimensional partition function, reviewed in appendix \ref{app:Z-3d},
we redefine the chemical potentials as
\begin{align}
  \label{eq:antonello-catthivo}
&
p = e^{2 \pi i r \omega_1}
&&
q = e^{2 \pi i r \omega_2}
&&
z = e^{2 \pi i r \sigma}
&&
s_a = e^{2 \pi i r m_a}
&&
t_a = e^{2 \pi i r \tilde m_a}
\end{align}
where $\sigma$, $m_a$ and $\tilde m_a$ parameterise the Cartan of the $U(N_c)$ gauge and the $SU(N_f)_L \times SU(N_f)_R$ flavour group respectively.  
In the limit $r \rightarrow 0$ the index reduces to the partition function.
In particular, the elliptic Gamma function $\Gamma_e$ reduces to the hyperbolic Gamma function $\Gamma_h$ $\eqref{eq:Gammahvbd}$ as \cite{Aharony:2013dha}
\begin{equation}
\label{eq:reduction}
\lim_{r\rightarrow 0} \Gamma_e \big( e^{2 \pi i r z} \big)=
e^{-\frac{i \pi}{6 \omega_1 \omega_2 r}(z-\omega)}
\Gamma_h(z)
\end{equation}
The divergent term in the RHS of (\ref{eq:reduction}) is proportional to the contribution of each multiplet to the gravitational anomaly.
Hence it will drop out when comparing the partition functions of two dual phases.
As discussed in \cite{Imamura:2011uw,Aharony:2013dha} also the four dimensional FI reduces to the three dimensional one.
With this prescription the index $\eqref{eq:index-el}$ reduces to the partition function for the three dimensional electric theory with the $\eta$ superpotential
$\eqref{eq:eta-superpot-adjoint}$
\begin{eqnarray}
\label{eq:Z-eta-el}
Z_{el}=W_{N_c,0}\big(\mu;\nu;\omega \Delta_X;\lambda\big)
\end{eqnarray}
where the function $W_{N_c,k}$ is defined in $\eqref{eq:Zdef}$
and $\Delta_X = 2/(k+1)$ denotes the $R$ charge of the adjoint.
From $\eqref{eq:antonello-catthivo}$ we obtain 
\begin{align} 
  \label{eq:real-masses-def}
  & \mu_a = \omega \, \Delta_Q + m_a && \nu_a = \omega \, \Delta_Q - \tilde m_a
\end{align} 
where in three dimensions we call the $R$ charge $R_Q = \Delta_Q$.
The balancing condition $\eqref{eq:bal-cond-4d}$ reduces to 
\begin{equation} \label{eq:bal-cond-3d} 
  \sum_{a=1}^{N_f+2} (\mu_a + \nu_a) = \omega (N_f +2 -N_c \, \Delta_X)
\end{equation} 
While in $4d$ the balancing condition reflects the anomaly cancellation for the $R$ symmetry, 
in $3d$ it is imposed by the superpotential $\eqref{eq:eta-superpot-adjoint}$.

In the magnetic phase the index reduces to the partition function 
\begin{align} 
\label{eq:Z-eta-mag}
Z_{mag} = \prod_{a,b=1}^{N_f+2}\prod_{j=0}^{k-1}
\Gamma_h\big(\mu_a+\nu_b+ j \omega \Delta_X\big) 
W_{\tilde N_c +2k,0}\big(\omega \Delta_X -\nu;\omega \Delta_X -\mu; \omega \Delta_X;-\lambda\big)
\end{align}
The four dimensional integral identity $I_{el} = I_{mag}$ reduces to $Z_{el} = Z_{mag}$,
with the constraint $\eqref{eq:bal-cond-3d}$.

\subsection{Flowing to Kim-Park duality}
Next we turn on the real masses in the electric theory as in $\eqref{eq:large-real-masses}$.
The parameters $\mu$ and $\nu$ become
\begin{align}
  \label{eq:large-real-masses-detailed}
&
\mu_a =
\left\{
\begin{array}{l}
m_a+m_A+\omega \Delta_Q \\
M-\frac{m_A N_f}{2}+\omega \Delta_{Q_M}\\
-M-\frac{m_A N_f}{2}+\omega \Delta_{Q_M} \\
\end{array}
\right.
&&
\quad \nu_a =
\left\{
\begin{array}{l}
-\tilde m_a + m_A+\omega \Delta_Q \\
-M-\frac{m_A N_f}{2 }+\omega \Delta_{Q_M}  \\
M-\frac{m_A N_f}{2}+\omega \Delta_{Q_M}  
\end{array}
\right.
&&
\begin{array}{r}
  \quad a=1,\dots,N_f \\
a=N_f+1 \\
a=N_f+2
\end{array}
\nonumber \\
\end{align}
Observe that the global $SU(N_f+2)^2$ symmetry is broken to $SU(N_f)^2
\times U(1)_A$ in the large $M$ limit. Indeed in this limit the $\eta$
superpotential that prevents the $U(1)_A$ symmetry disappears \cite{Aharony:2013dha}. 
At large $M$ (\ref{eq:Z-eta-el}) becomes 
\begin{equation}
  \label{eq:Zel-largeM}
  Z_{el} = e^{-\frac{i \pi}{2 \omega_1 \omega_2} (4 M N_c (m_A N_f-2 \omega  (\Delta_{Q_M}-1)))}
  W_{N_c,0}\big(\mu;\nu;\omega \Delta_X;\lambda\big)
\end{equation}
In the magnetic case the situation is more involved.
The real masses can be read from the electric theory.
However, in order to reproduce the divergent prefactor in $\eqref{eq:Zel-largeM}$ also on the dual side, 
we need to turn on a non-trivial vacuum structure for the scalar $\sigma$ in the vector multiplet
\begin{equation}
  \label{eq:vacuum-large-M-mag}
\sigma_i=
\left\{
\begin{array}{rl}
0&~~~~~~i=0,\dots,k N_f-N_c \\
M&~~~~~~i=kN_f-N_c+1,\dots,k(N_f+1)-N_c\\
-M&~~~~~~i=k(N_f+1)-N_c+1,\dots,k(N_f+2)-N_c
\end{array}
\right.
\end{equation}
This is consistent with the vaccum structure choosen in \cite{Nii:2014jsa,Aharony:2013dha}.
Note that in absence of the $R$ symmetry breaking polynomial in the adjoint the gauge symmetry $U(\tilde N_c+2k)$ is broken to $U(\tilde N_c) \times U(k)^2$.
In the large $M$ limit (\ref{eq:Z-eta-mag}) becomes 
\begin{equation} 
  \begin{aligned}
     Z_{mag} & = \,  e^{-\frac{i \pi}{2 \omega_1 \omega_2} \left(4 M N_c (m_A N_f-2 \omega  (\Delta_{Q_M}-1))\right)}
\prod_{j=0}^{k-1} \prod_{a,b=1}^{N_f}
    \Gamma_h\left(\mu_a+\nu_b+ j \, \omega \Delta_X  \right)  \\
& \times \, 
W_{k N_f  -N_c,0}\left(\omega \Delta_X -\nu_a;\omega \Delta_X-\mu_a; 
\omega \Delta_X;-\lambda\right) 
\, \; Z_+ \;  Z_-
  \end{aligned}
\end{equation}
The additional terms $Z_{\pm}$ are the partition functions of the two $U(k)$ sectors
\begin{equation}
Z_{\pm}=\prod_{j=0}^{k-1} \Gamma_h(m_{j, \pm}) \; W_{k,0}
\left(\mu_{\pm},\nu_{\pm},  \omega \Delta_X, \lambda_{\pm}\right)
\end{equation}
with
\begin{align}
 m_{j,\pm} = - m_A N_f+ 2 \omega  \Delta_{Q_M} + j \, \omega \Delta_X  \, ,
&&
\mu_{\pm}=\nu_{\pm} = \frac12 m_A N_f+\omega\left(\Delta_X-\Delta_{Q_M}\right)
\end{align}
and effective FI terms 
\begin{equation}
  \lambda_\pm = -\lambda \pm  
  \left(N_f \, m_A + \omega  \left(2 N_f \Delta_Q + 2 \Delta_{Q_M} - (N_f +1) (k-1) \Delta_X \right)\right) 
\end{equation}
Recall that the $U(k)^2$ sector is the enhancement of the $k$ $U(1)^2$ sectors discussed in \cite{Nii:2014jsa}.
As we mentioned at the end of section \ref{sec:4d3d} the $U(k)^2$ sector can be dualised to a set of $6k$ singlets. 
We can see this on the partition function exploiting the integral identity \cite{VdB}
\begin{equation}
\label{eq:IdA.5}
W_{N_c,0}(\mu;\nu;\omega \Delta_X;\lambda) \!= \!
\prod_{j=0}^{N-1}
\!\Gamma_h  \Big(\omega -\frac{\mu+\nu}{2} -j \, \omega \Delta_X\pm \frac{\lambda}{2}\Big) \,
\Gamma_h(\mu+\nu + j \,  \omega \Delta_X )
\end{equation}
$4k$ singlets acquire mass from a superpotential and are integrated out.
On the partition function this is reflected in the relation $\Gamma_h (z) \Gamma_h(2\omega-z)=1$.
In this process we used the constraint on the $R$ charges coming from $\eqref{eq:bal-cond-3d}$
\begin{equation}
\label{eq:bal-cond-3d-dual}
N_c \Delta_X +N_f \left(\Delta_Q-1\right)+2 \left(\Delta_{Q_M}-1\right)=0
\end{equation}
We are left with
\begin{equation}
Z_+ Z_- = \prod_{j=0}^{k-1} \Gamma_h \Big(
\pm\frac{\lambda }{2}-m_A N_f+\omega  \left(\left(j-N_c+1\right) \Delta_X +N_f \left(1-\Delta _Q\right)\right)
\Big)
\end{equation}
This is the contribution to the partition function from the $2k$ singlets which remain light.
They have exactly the right global charges for coupling through the superpotential interaction $\eqref{eq:W-KP-mag}$ with the magnetic monopoles $\tilde t_{i,\pm}$.
It is hence natural to identify them with the electric monopoles $t_{i,\pm}$.
Eventually we arrive at the identity
\begin{equation} \label{eq:parkPF}
  \begin{aligned}
W_{N_c,0}\!&\left(\mu;\nu;\omega \Delta_X;\lambda\right)
=
\prod_{j=0}^{k-1} \prod_{a,b=1}^{N_f} \Gamma_h(\mu_a+\nu_b+ j \, \omega \Delta_X) \\
\times &
\, \prod_{j=0}^{k-1} \Gamma_h \Big(
\pm\frac{\lambda }{2}-m_A N_f+\omega  \left(\left(j-N_c+1\right) \Delta_X +N_f \left(1-\Delta _Q\right)\right)
\Big)
\\
\times 
& \, \,
W_{k N_f-N_c,0}\left(\omega \Delta_X-\nu;\omega \Delta_X-\mu;\omega \Delta_X;-\lambda\right)
  \end{aligned}
\end{equation}
Note that as expected this identity holds without any balancing condition.
In the limiting case $N_f=1$ and $k=N_c$ it corresponds to the identity $\eqref{eq:IdA.5}$ from \cite{VdB}.
This fact is as a consistency check of the procedure.

\subsection{Flowing to the case with CS terms}

In \cite{Niarchos:2008jb,Niarchos:2009aa} another three dimensional duality with adjoint fields has been discussed. 
It has a CS gauge interaction and extends the duality \cite{Giveon:2008zn}. 
The electric phase is a $U(N_c)_K$ gauge theory at CS level $K$ with $N_f$
(anti-)fundamentals and one adjoint, with superpotential $\eqref{eq:W-el-adj}$.
The magnetic phase is a $U(k(N_f+|K|)-N_c)_{-K}$ gauge theory at CS level $-K$ with $N_f$ (anti-)fundamentals, one adjoint and $k$ singlets,
with superpotential (\ref{magn4d}).

The duality can be derived from Kim-Park duality by an RG flow.
In the electric phase we consider a $U(N_c)_0$ gauge group and $N_f+K$ (anti-)fundamentals, giving a large real mass to $K$ of them.
In the dual $U(k(N_f+K)-N_c)$ theory the fields acquire their masses accordingly. 
More explicitly, $K$ magnetic flavours $(q, \tilde q)$, the electric monopoles $t_{j, \pm}$ and $K^2+2K N_f$ components of each meson $M_j$ become heavy.
In the large mass limit these fields are integrated out. 
This procedure generates the CS levels and one recovers the duality of \cite{Niarchos:2008jb,Niarchos:2009aa}.

One can study this RG flow at the level of the partition function.
Assigning the masses as described above and taking the large mass limit, the identity $\eqref{eq:parkPF}$ becomes\footnote{Here
we choose without loss of generality positive real masses for the electric quarks.}
\begin{equation}
  \label{eq:identity-Niarchos}
  \begin{aligned}
W_{N_c,K}\left(\mu;\nu;\omega \Delta_X;\lambda\right) 
= & \, e^{\frac{i \pi}{2 \omega_1 \omega_2} \, \phi} \zeta^{-k(2+K^2)} \; \,
\prod_{j=0}^{k-1}\prod_{a,b=1}^{N_f} \!  \left(\mu_a+\nu_b+j \, \omega \Delta_X \right) \\
\times \, & W_{k(N_f+K)-N_c,-K}\left(\omega \Delta_X -\nu;\omega \Delta_X-\mu;\omega \Delta_X;-\lambda\right) 
  \end{aligned}
\end{equation}
where $\zeta=\exp\big(\frac{\pi i (\omega_1^2 +\omega_2^2)}{24 \omega_1 \omega_2}\big)$ and the extra phase is
\begin{equation} \label{eq:phase} 
  \begin{aligned}
\phi&=
\omega  m_A \big(2 k N_f \left(2(K- N_f) \Delta _Q- 2 N_c \Delta_X  - 2 N_f  +K (k-1)\Delta_X \right)\big)
 \\
&+2 k m_A^2 N_f \left(K-N_f\right)-\frac{k \lambda ^2}{2}+
k K \sum _{a=1}^{N_f} \left(\mu _a^2+\nu _a^2\right)
 \\
&-
k \omega^2 \Big( \Big(2 N_c \left(N_c+(k+1) N_f \left(\Delta _Q-1\right)-k K\right)+\frac{k^2 \left(11 K^2+2\right)+K^2-2}{12}\Big) \Delta_X^2
 \\
&-
2 N_f \Big(K \big(\Delta _Q^2+(k-1) \Delta _Q \Delta_X +\frac53 (\Delta_X-1) -1 \big)- N_f \left(\Delta _Q-1\right)^2\Big) \Big)
  \end{aligned}
\end{equation}
Equation $\eqref{eq:identity-Niarchos}$ is the identity between the electric and the magnetic partition functions of the duality in
\cite{Niarchos:2008jb,Niarchos:2009aa}.
Note that the limiting case $N_f=0$, $K=1$ and $k=N_c$ is the third integral identity in theorem $5.6.8$ of \cite{VdB}.
Similar dualities appeared in the physics literature \cite{Jafferis:2011ns,Kapustin:2011vz,Agarwal:2012wd,Imamura:2012rq}.

\subsubsection*{Comments on the phases and CS terms}

In this section we studied RG flows by turning on real masses.
These flows generate not only CS terms for the gauge but also for the global symmetries.
The real masses arise as vevs after weakly gauging the global symmetries and the global CS give rise to a phase, quadratic in $\omega$ and in the mass parameters
\cite{Benini:2011mf}.
When flowing from the duality with $\eta$ superpotential to Kim-Park duality, we checked that the generated CS all cancel.
In fact, in the relation \ref{eq:parkPF} there is no extra phase.
However in the flow to the duality of \cite{Niarchos:2008jb,Niarchos:2009aa} we found a global CS action.
We checked that it precisely corresponds to the phase (\ref{eq:phase}).

\section{Conclusions and further directions}
\label{sec:conclusioni}
In this paper we studied the dimensional reduction of four dimensional dualities to three dimensions.
We considered the reduction of KSS duality discussed in \cite{Nii:2014jsa}.
This yields a new $3d$ duality with the superpotential $\eqref{eq:eta-superpot-adjoint}$.
Other previously studied $3d$ dualities can be deduced from this by an RG flow.
We studied how the identity of the $4d$ superconformal index leads to identities between $3d$ partition functions.
As a partial check of these identities we observed their agreement with theorem $5.6.8$ of \cite{VdB}.
Furthermore we matched the phases in the integral identities with the global CS terms generated along the RG flows.

Let us comment on the validity of the mathematical identities of section \ref{sec:I-mathcal-Z}.
The parent relation in four dimensions $I_{el}=I_{mag}$ has not been rigorously proven in the mathematical literature,
though several partial checks have been performed (see \cite{Spiridonov:2009za} for references). 
Our derivation of the identities between hyperbolic hypergeometric integrals 
$Z_{el}=Z_{mag}$ in section \ref{sec:I-mathcal-Z} is not a proper mathematical derivation.
However we can think of this as an instance where physical intuition may be helpful in making predictions about unknown mathematical results.

We conclude with an outlook of some possible lines of future research.
Note that the relation between the partition functions of the $SU(N)$ duality \cite{Park:2013wta} can be found by the 
``ungauging'' procedure of \cite{Aharony:2013dha}.
Alternatively one can apply the reduction discussed in \cite{Nii:2014jsa} to the $4d$ superconformal index.
It would be desirable to extend the dualities studied in this paper by considering a chiral flavour sector,
as done for the case without adjoint in \cite{Benini:2011mf}.
This would also be useful in studying the inverse flow from the dualities with CS terms in \cite{Niarchos:2008jb,Niarchos:2009aa} to Kim-Park duality, 
along the lines of \cite{Intriligator:2013lca,Khan:2013bba,Amariti:2013qea}.
On the partition function this corresponds to deriving the identity $\eqref{eq:parkPF}$ from $\eqref{eq:identity-Niarchos}$.
Moreover, one might want to reduce the $4d$ dualities with multiple adjoint matter \cite{Brodie:1996vx} to $3d$,
obtaining additional integral identities. 
Another interesting extension of the programme is the analysis of dualities with real gauge groups and tensor matter.

\section*{Acknowledgements}

A.A. is supported by the Institut de Physique Th\'eorique Philippe Meyer at the
\'Ecole Normale Sup\'erieure.
C.K. acknowledges support by ANR grant 12-BS05-003-01.

\appendix

\appendix

\section{The three dimensional partition function}
\label{app:Z-3d}

The partition function on  the squashed $S^3$ for a gauge group $U(N_c)_K$  
is a matrix integral over the Cartan of the gauge group,
parametrised by the scalar $\sigma$ in the $\mathcal{N}=2$ vector multiplet \cite{Jafferis:2010un,Hama:2010av,Hama:2011ea}.
For $N_f$ (anti-)fundamental flavours and one adjoint the matrix integral is
\begin{eqnarray}
\label{eq:Zdef} 
W_{N_c,K} \big( \mu; \nu ; \tau; \lambda \big) = 
&&\frac{\Gamma_h(\tau)^{N_c}}{N_c !} \int \prod_{i=1}^{N_c} d \sigma_i \;
 e^{\frac{i \pi }{2\omega_1 \omega_2} (2 \lambda \tr \sigma - 2 K \tr \sigma^2)}\;
 \prod_{1 \le i < j \le N_c } \!\! \frac{\Gamma_h(\tau \pm (\sigma_i - \sigma_j))}{\Gamma_h(\pm(\sigma_i -\sigma_j))} \; \nonumber \\
 \times &&  \prod_{i=1}^{N_c} \prod_{a,b=1}^{N_f} \Gamma_h(\mu_a + \sigma_i) \Gamma_h(\nu_b -\sigma_i) \;
\end{eqnarray}
where $\omega_1$ and $\omega_2$ are related to the real squashing parameter $b$ of $S^3$ as $\omega_1 = i b$ and $\omega_2 = i / b$,
and $\omega \equiv (\omega_1 + \omega_2)/2$.
The function $\Gamma_h$ is the hyperbolic Gamma function, it can be written as \cite{VdB}
\begin{equation}
\label{eq:Gammahvbd}
\Gamma_ h(z;\omega_1,\omega_2) \equiv \Gamma_h(z) \equiv
\prod_{n,m=1}^{\infty}
\frac{(n+1)\omega_1+(m+1) \omega_2-z }{n \omega_1+m \omega_2+z}
\end{equation}
We also introduced the shorthand notation $\Gamma_h(\pm x) \equiv \Gamma_h(x)\Gamma_h(-x)$.
Let us comment on the different terms in the partition function and their physical interpretation. 
The exponential is the contribution from the CS action and the FI term $\xi = 2 \lambda$.
The hyperbolic Gamma functions are the $1$-loop contributions of the various multiplets,
their arguments reflect the charges under the local and global symmetries.
More precisely, $\pm(\sigma_i -\sigma_j)$ are the weights of the adjoint representation while $\pm \sigma_i$ the ones of the (anti-)fundamental.
Similarly $\tau, \mu$ and $\nu$ are collective parameters for the weights under the global symmetries
for the adjoint, the fundamental and the anti-fundamental respectively.

\section{Some comments on the infinite mass limit of the partition function}
In this appendix we discuss some issues concerning the large $M$ limit on the partition function.
As stated in section \ref{sec:I-mathcal-Z} when taking the large mass limit $\eqref{eq:large-real-masses-detailed}$ 
one needs to choose the non-trivial vacuum $\eqref{eq:vacuum-large-M-mag}$ in the magnetic theory in order to preserve the duality.
The necessity, in order to preserve the duality,
of different vacua in such real mass flows is well known in field theory \cite{Intriligator:2013lca,Aharony:2013dha,Nii:2014jsa},
here we want to discuss it at the level of the partition function.
We repeat the analogue discussion for SQCD in \cite{Aharony:2013dha}, including also adjoint matter fields.

Recall that we start from the identity $Z_{el} = Z_{mag}$ between the formulae $\eqref{eq:Z-eta-el}$ and $\eqref{eq:Z-eta-mag}$,
which is valid whenever the balancing condition $\eqref{eq:bal-cond-3d}$ is satisfied.
The integrals are over the Cartan of the gauge group and match as functions of the complex parameters\footnote{Recall
that $\mu$ and $\nu$ are holomorphic combination of the real masses and the $R$-charges, see equations $\eqref{eq:real-masses-def}$.}
$\mu$ and $\nu$,
the integer $k$ and the real FI parameter $\lambda$.
The identity descends directly from the identity between the corresponding $4$d indices
and reflects the duality of effective $3$d theories on the circle. 

When parametrising the real masses as in $\eqref{eq:large-real-masses-detailed}$ and $\eqref{eq:vacuum-large-M-mag}$,
the integrals still coincide for any \emph{finite} value of $M$,
taking $\sigma \rightarrow  \sigma + M$ is just a shift in the integration variables.
However, in the limit $M \rightarrow \infty$ the integrals are divergent.
Their leading contribution for large $M$ is approximated by a saddle point\footnote{We
assume the generic case of non-degenerate saddle points. 
}.
Shifts in $\sigma$ of order $M$ correspond to picking up saddle points from regions at order-$M$-distance in the Coulomb branch.

Let us use the field theory intuition to associate saddle points to the vacua \cite{Intriligator:2013lca} in the massive theory.
At these points some charged fields remain light and give a finite contribution to the integral.
In general, there are more than one such saddle points, we want to pick the leading one.
Which saddle point gives the leading contribution to the integral depends on the range of parameters.
From the equations of motion we see that for real masses $\eqref{eq:large-real-masses-detailed}$ we
can have the two vacua  $\sigma = 0$ and the Higgsed $\eqref{eq:vacuum-large-M-mag}$.
We have checked that for an $R$-charge of the quarks $\Delta_Q$ smaller then the critical $R$-charge 
\begin{align} 
  \label{eq:crit-R-charge}
  \Delta_0 = 1 - \frac{N_c -k}{N_f} \Delta_X
\end{align} 
the saddle point at the vacuum $\sigma = 0$ gives the leading contribution to the electric integral $\eqref{eq:Z-eta-el}$,
while for $\Delta_Q > \Delta_0$ the leading contribution comes from the Higgsed vacua $\eqref{eq:vacuum-large-M-mag}$.
The leading large-$M$ contribution in a trivial vacuum on the electric side $\eqref{eq:Z-eta-el}$
coincides with a Higgsed vacuum on the magnetic side $\eqref{eq:Z-eta-mag}$ (where also the contribution from the mesons is included).\footnote{For
pure SQCD the same mapping of the vacua has been shown in \cite{Niarchos:2012ah} by analysing the parameter range of validity of the partition function.
}
In particular, also for the magnetic integral the critical $R$-charge is $\eqref{eq:crit-R-charge}$,
but here the saddle point in the trivial vacuum dominates for $\Delta_Q > \Delta_0$ while the one in the Higgsed vacuum dominates for $\Delta_Q < \Delta_0$.

However, both choices $\Delta_Q > \Delta_0$ and $\Delta_Q < \Delta_0$,
or, correspondingly, the two choices of (mutually different) vacua for the electric and the magnetic side yield the same pure $3$d relation,
one valid for $\Delta_Q < \Delta_0$ and one for $\Delta_Q > \Delta_0 - 2k \Delta_X/N_f$. 
This covers the whole range of $\Delta_Q$ and we conclude the general validity of $\eqref{eq:parkPF}$.

Let us finish with a comment.
Looking at the partition functions one might wonder how to discriminate between the magnetic the electric side.
Indeed they differ only by the appearance of the mesons, which can be carried from one to the other side using the identity 
\begin{align} 
\label{eq:Wmass}
  \Gamma_h(2 \omega - z) \Gamma_h (z) =1
\end{align} 
Multiplying both sides of the integral identity between $\eqref{eq:Z-eta-el}$ and $\eqref{eq:Z-eta-mag}$ with
\begin{align} 
\label{eq:mesonN}
  \prod_{a,b=1}^{N_f} \prod_{j=0}^{k-1} \Gamma_h (2 \omega - (\mu_a + \nu_b + (k+1-j) \omega \Delta_X ))
\end{align} 
and using
\begin{align} 
  \prod_{a,b=1}^{N_f} \prod_{j=0}^{k-1} 
  \Gamma_h (\mu_a + \nu_b + j \omega \Delta_X ) \Gamma_h (2 \omega - (\mu_a + \nu_b + (k+1-j) \omega \Delta_X )) =1
\end{align} 
we obtain the same integral identity with the roles of the electric and magnetic sides interchanged.
\emph{I.e.\ }the mesons are on the side which was the electric integral before and 
we re-obtain the identity between $\eqref{eq:Z-eta-el}$ and $\eqref{eq:Z-eta-mag}$ where now
$N_c \rightarrow M_c = \widetilde N_c +2k$ and $\widetilde N_c \rightarrow \widetilde M_c  = k N_f - M_c$.

We can rephrase this ambiguity in field theory language.
Multiplying  the term (\ref{eq:mesonN}) in the magnetic side of the partition function 
corresponds to add, to the magnetic superpotential, the contribution of $k$ extra singlets $N_j$, constrained by an
interaction of the form
\begin{align} 
\label{eq:Wextramagn}
  \Delta W_{mag} = \sum_{j=0}^{k-1} M_j N_{k+1-j}
\end{align} 
Integrating out these massive fields corresponds to using the relation (\ref{eq:Wmass}).
The term (\ref{eq:Wextramagn}) is dual to
\begin{align} 
  \label{eq:W-el-N}
  \Delta W_{el} = \sum_{j=0}^{k-1} N_j Q X^{k+1-j} \tilde Q  
\end{align} 
where $N_j$ is identified with $q Y^{j} \tilde q$.
This extra term corresponds to the contribution of the term (\ref{eq:mesonN})
in the electric partition function.
The superpotential $\eqref{eq:W-el-N}$ only constraints the charges of the fields $N_j$.

The net effect of this transformation has been to ``move'' the mesons from the 
RHS to the LHS of the equality between the partition functions. However, this operation does not change the 
ranks of the gauge and flavor groups involved in the duality. 
Also the scaling properties of the matrix integrals, 
for a given choice of $R$-charges, has not been modified.
It follows that the same vacua in the electric and in the magnetic case have to be kept,
even when the mesons are moved on the electric side.

\bibliographystyle{JHEP}
\bibliography{BibFile}

\end{document}